\documentstyle[aps,prl,epsf]{revtex}

\newcommand{\bc}{\begin{center}}
\newcommand{\ec}{\end{center}}
\newcommand{\bt}{\begin{tabular}}
\newcommand{\et}{\end{tabular}}
\newcommand{\bdes}{\begin{description}}
\newcommand{\edes}{\end{description}}

\newcommand{\be}{\begin{equation}}
\newcommand{\ee}{\end{equation}}
\newcommand{\bea}{\begin{eqnarray}}
\newcommand{\eea}{\end{eqnarray}}
\newcommand{\non}{\nonumber}

\newcommand{\ba}{\begin{array}}
\newcommand{\ea}{\end{array}}

\newcommand{\bff} { {\bf f}}

\newcommand{\br} { {\bf r}}

\begin{document}
\advance\textheight by 0.5in
\advance\topmargin by -0.2in
\draft

\twocolumn[\hsize\textwidth\columnwidth\hsize\csname@twocolumnfalse%
\endcsname

\title{Quasicontinuum Models of Interfacial Structure and Deformation}

\author{V. B. Shenoy$^1$, R. Miller$^1$,  E. B. Tadmor$^2$, 
R. Phillips$^1$ and M. Ortiz$^3$}

\address{1) Division of Engineering, Brown University, Providence, RI 02912}

\address{2) Department of Physics, Harvard University, Cambridge, MA 02138}

\address{3) Graduate Aeronautical Laboratories, California Institute of Technology, Pasadena, CA 91125}

\date{\today}

\maketitle

\begin{abstract}
Microscopic models of the interaction between grain boundaries (GBs)
and both dislocations and cracks are of  importance in
understanding the role of microstructure in altering the mechanical
properties of a material.  A recently developed mixed atomistic
and continuum method is extended to examine the interaction between
GBs, dislocations and cracks.  These calculations
elucidate plausible microscopic mechanisms for these defect interactions 
and allow for the quantitative evaluation of critical parameters such
as the stress to nucleate a dislocation at a step on a GB and
the force needed to induce GB migration.
\end{abstract}
]


With the continuing development of more accurate, less expensive
models for atomistic interactions and expansion of computational
resources, there is growing interest in the modeling of materials from
fundamental principles rather than phenomenological approaches.  An
outstanding problem in this regard is the role of microstructure in
determining material properties.  The influence of microstructure
(e.g. grain size and shape) on the mechanical properties of materials
is clearly revealed, for example, in the yield strength and the
fracture toughness \cite{lawn}.  A first step in the microscopic
determination of the role of microstructure in governing such
properties is the elucidation of plausible mechanisms whereby
dislocations and cracks, the primary agents of permanent deformation,
interact with the boundaries that make up that microstructure.  One of
the key challenges posed by such calculations is the simultaneous
operation of multiple scales in the problem.  In this letter, we
present a recently developed model for treating multiple scales.  We
then demonstrate the application of the model to three examples: the
deformation of a stepped grain boundary (GB), the interaction of
lattice dislocations with GBs and the interaction of cracks with GBs.

Recently, the quasicontinuum method was proposed to allow for a
seamless treatment of multiple length scales \cite{tadmora,tadmorb}. This
mixed atomistic-continuum formulation is based on a finite element
discretization of a continuum mechanics variational principle.  The
finite element method serves as the numerical engine for determining
the energy minimizing displacement fields, while atomistic analysis is
used to determine the energy of a given configuration. This is in
contrast to standard finite element approaches where the constitutive
input is made via phenomenological models. The method is successful in
capturing the structure and energetics of dislocations.

In this paper we generalize the method to allow for the treatment
of interfaces, and show how the formulation allows for the
simultaneous treatment of dislocations, material interfaces and
cracks.  We begin with the recognition that from the microscopic
perspective the body may be regarded as a collection of $N$ atoms, the
total potential energy of which is given by
\vspace{-0.1truein}
\be
\Pi = \sum_{i= 1}^{N} E_i({\bf r}_1,...,{\bf r}_N) - \sum_{i=1}^N {\bf
f}_i \cdot {\bf r}_i \nonumber 
\ee
where ${\bf r}_i$ is the position of the atom $i$,
${\bf f}_i$ is the
external force on that atom, and $E_i$ is its energy
 as would be computed from an atomistic model such as the embedded
atom method\cite{dawbask} used here.   One of the primary objectives
in the formulation of the method is to eliminate the redundant
atomistic degrees of freedom associated with the regions
of the body far from extended defects and hence subject to
displacement fields which are slowly varying on the atomic scale.
To achieve the requisite degree of
freedom reduction we select $M$ {\em representative atoms} from the
$N$ atoms $(M << N)$, chosen to best represent the energetics of
the body, the positions ${\bf r}_\alpha (\alpha = 1,...,M)$ of which serve
as the reduced set of  degrees of freedom.
The body is now divided into disjoint cells  such
that each cell contains exactly one representative atom. The key
energetic approximation is that the energy of all of the atoms in a 
given cell is the same as that of the cell's representative
atom. The
positions of the atoms  that are not treated explicitly
are  obtained by  interpolating
the nodal values of the displacements  using a finite
element  mesh which is constructed with the
representative atoms as the nodal points. (One possible implementation
of this strategy in two dimensions 
 is to use the Voronoi polygons\cite{sugihara} surrounding the 
representative atoms as the cells, and  the geometric dual of the
Voronoi tiling, the Delaunay triangulation\cite{sloan}, as the finite
element mesh.) 

Given the scheme described above, the approximate potential energy depends only on
the positions of the representative atoms ${\bf r}_\alpha$ and can be
written as
\vspace{-0.1truein} \be \Pi_{reduced} = \sum_{\alpha= 1}^{M} n_\alpha
\bar{E}_\alpha(\br_1,...,\br_M) - \sum_{\alpha=1}^M n_\alpha
\bar{\bff}_\alpha \cdot \br_\alpha, \non \ee where $n_\alpha$ is the
number of atoms represented by atom $\alpha$, $\bar{E}_\alpha$ is the
energy of that representative atom and $\bar{\bf f}_\alpha$ is the
effective force acting on the $\alpha^{th}$ representative atom. In
practice, $\bar{E}_\alpha$ is computed in two different ways.  When
the representative atom is located in a region undergoing strongly
non-uniform deformation, $\bar{E}_\alpha$ is computed using the usual
atomistic rule in which a given atom is surrounded by its complement
of neighbors and the resulting energy per atom is computed. On the
other hand, if the representative atom experiences a slowly varying
deformation the energy is still computed atomistically, but with the
assumption that its environment is distorted according to the local
gradients of deformation.  The details of the criteria that dictate
which scheme is used will be described elsewhere.  The key outcome
gained in the implementation of the strategy described above is the
incorporation of the relevant atomistic nonlinearity and nonlocality
that allows for the emergence of defects such as dislocations and
cracks, without the attendant singularities that plague linear elastic
analyses, or the burden of many redundant atomistic degrees of freedom.

As in our earlier work with this method, a key criterion to the
validity of the model is how well the results of conventional
atomistics are recovered in the context of defects of known structure.
For the purposes of the present paper, the method must successfully
reproduce the known static geometric structures of GBs.
As a test of the method, we have examined the structure of a range of
GBs in several fcc metals.  For the moment, we have
confined our attention to symmetric tilt boundaries, using
embedded-atom type potentials.  The key quantitative tests of the
outcome of these calculations are an appropriate reckoning of i) the
interfacial energy, and ii) the interfacial structure.  An indication
of the typical energy differences between the quasicontinuum result
and the associated direct atomistic calculation is demonstrated by a
$\Sigma 5(210)$ GB in Au where the energy as obtained by
conventional atomistics (676$mJ/m^2$) \cite{aukland} and the method
described here (670$mJ/m^2$) are in close agreement.  Similarly, in
all of the cases we have considered (i.e.  $\Sigma 5(210)$ in Al
and Cu, $\Sigma 3(111)$ in Al, $\Sigma 99(557)$ in Al, and $\Sigma
21(\bar{2}41)$ in Al, Au and Ni), the atomic level geometry at the
interface obtained using the quasicontinuum method advocated here was
for practical purposes identical to that obtained using direct
atomistic simulation.  Consequently, it is of interest to turn the
method to the analysis of interfacial deformation.

\begin{figure}
\narrowtext \epsfxsize=3.4truein \vbox{\hskip 0.15truein
\epsfbox{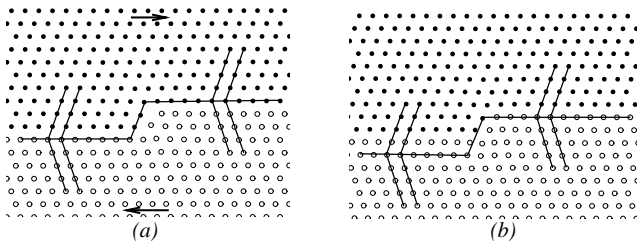}}
\medskip
\caption{Step motion under applied shear stress.  Atomic positions
associated with stepped twin boundary, illustrating the motion
of the boundary perpendicular to the boundary plane as
a result of an applied shear stress. (a) Initial configuration.  (b)
Final configuration after the application of the critical stress.}
\end{figure}

As an example of such deformation, we consider a stepped twin boundary
($\Sigma 3 (111)$) in Al that is subjected to a remotely applied shear
stress.  It is found that once a certain critical shear stress
($\approx 0.033\mu$, where $\mu$ is the shear modulus relevant to the
orientation of interest here) is attained, the step on the 
boundary serves as a source for ${a_0 \over 6}[\bar{1} \bar{1} 2]$
dislocations which are swept out along the boundary plane.  The
consequence of the passing of these dislocations is the net downward
motion of the twin boundary shown in fig. 1.  One question of
importance concerning the type of simulations described above is that
of the convergence of the results with respect to both system size and
mesh geometry.  We have found that altering the system size reveals
that the critical stress of $\approx 0.033\mu$ necessary to move the
boundary is reproduced to within $\pm 0.003\mu$ from one simulation to
the next.  It is interesting to contrast this stress with a typical
Peierls stress for a straight dislocation.  For example, in the case
of a screw dislocation in this metal, we have found \cite{shenoy} that
the Peierls stress is 0.00068$\mu$, nearly fifty times smaller than
the critical stress for advancing the twin boundary.  As another
comparison to set the scale of the stresses determined here, the
stress to induce motion of the twin boundary can be compared with that
to operate a Frank-Read source which is $\sigma \approx \mu b / L$,
where $L$ is the width of the source \cite{hullbacon}.  In light of
this estimate, the stress to induce motion of the twin boundary is of
the same order as that to operate a Frank-Read source of width
$\approx 35b$ (where $b$ is a typical Burgers vector). Although
typical Frank-Read sources are larger than $35b$ and hence operate at
even lower stresses, the stress found to stimulate motion of the twin
boundary is still significantly smaller than the ideal shear strength,
and is another example of the ``lubricating'' effect of
heterogeneities in the motion of extended defects.

Despite the existence of useful continuum models of
dislocation-GB interactions, it remains a crucial
challenge to uncover the microscopic processes that transpire once
the dislocation core is in the proximity of a GB.  Our
earlier work on simulating nanoindentation \cite{tadmorb} suggests
the possibility of using nanoindentation induced dislocations to
probe the interaction between dislocations and a GB.  As
a model system, we consider a block oriented such that (111) planes
are positioned to allow for the emergence of dislocations which then
travel to the $\Sigma 21(\bar{2}41)$ GB which waits
approximately 200$\AA$ beneath the surface (c.f. figure 2a).

Because of the relatively high stacking fault energy associated with
the EAM potentials for Al we used \cite{ercadams}, the dislocations
nucleated at the free surface as a result of the indentation process
are produced as rather closely spaced ($15\AA$) Shockley partials.  As
seen in the left frame in part (b) of fig. 2, the Shockley partials
have been absorbed at the GB with the creation of a step at the
GB. This geometry can be
rationalized on the basis of the underlying displacement shift
complete (DSC) lattice \cite{king} associated with this
symmetric GB.  We find that the lattice dislocation ${a_0 \over
2} [\bar{1} 1 0]$ can be split into two DSC lattice vectors,
\begin{equation}
\frac{a_o}{2}[\bar{1} 1 0] = \underbrace{\frac{a_o}{14}[\bar{3} \bar{1} \bar{2}]}_{GB Dislocation}
+ \underbrace{\frac{a_o}{7} [\bar{2} 4 1]}_{Step}, 
\end{equation}
where ${a_0 \over 14}[\bar{3} \bar{1} \bar{2}]$ is the Burgers
vector of a GB dislocation parallel to the GB
and ${a_0 \over 7}[\bar{2} 4 1]$ is the  vector associated with the step.

As the load is increased, a second pair of Shockley partials is
nucleated and they are not immediately absorbed into the
GB and consequently form a pile-up (c.f. fig. 2).  These
dislocations are not absorbed until a much higher load
level is attained.  Due to the lack of dislocation activity
in the neighboring grain, it may be concluded that this GB
does not aid slip transmission.  

\begin{figure}
\narrowtext \epsfxsize=3.4truein \vbox{\hskip 0.15truein
\epsfbox{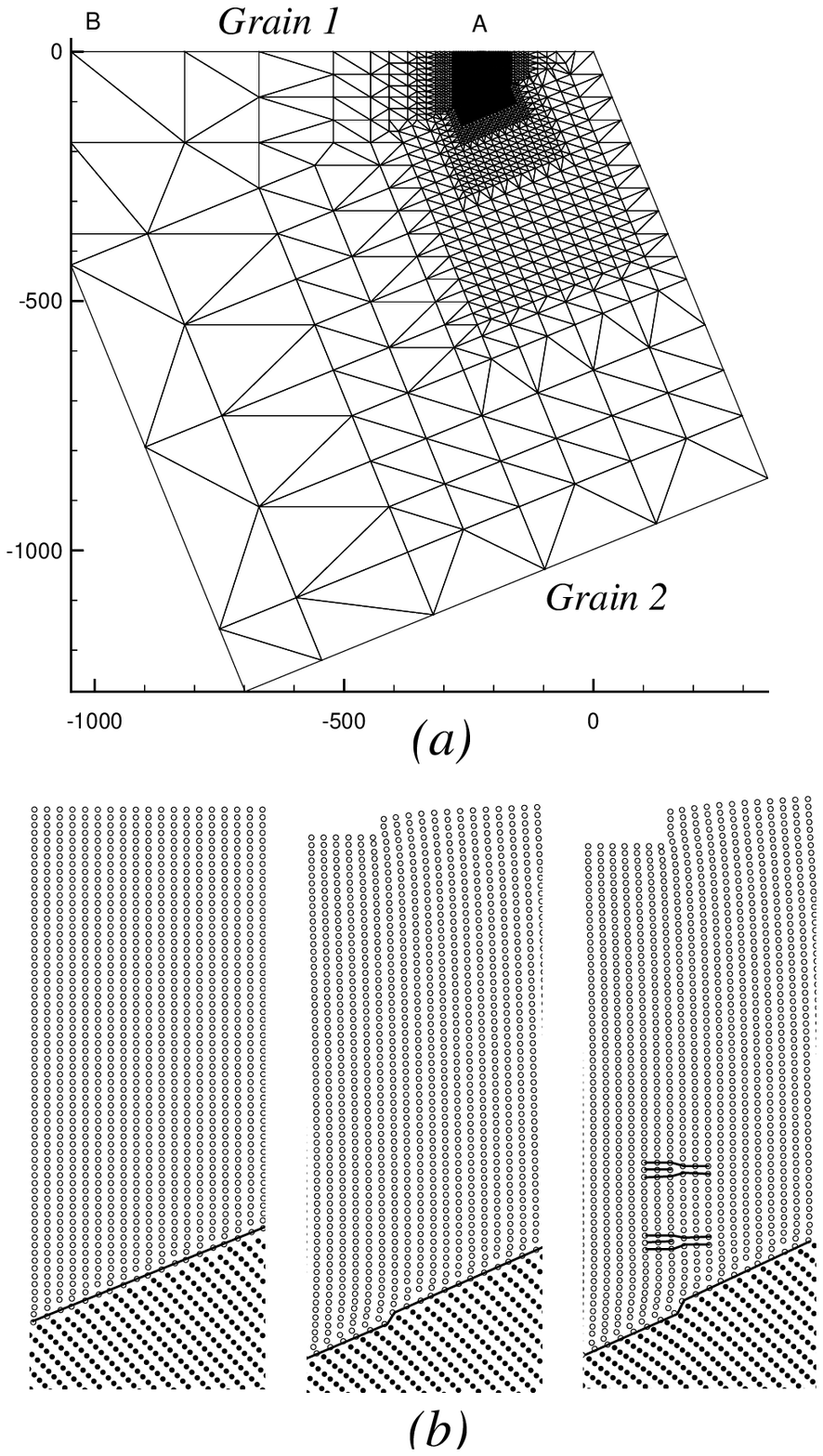}}
\medskip
\caption{(a) Finite element mesh used to model dislocation/GB
interaction. The surface marked $AB$ is rigidly indented in order to
generate dislocations at $A$.  (b) Snapshots of atomic positions at
different stages in the deformation history.  Absorption of the first
pair of dislocations at the GB results in a step, while the second
pair form a pile-up.}
\end{figure}

As a final example of the synthetic view of extended defects
afforded by this method, we consider the interaction
between a brittle crack and a GB.
The interaction of cracks and interfaces  pose a variety of challenging and
important problems.  One issue that can be considered within the confines
of the method presented here is that of the interaction of a crack propagating
by  cleavage as it impinges upon a GB
in its path.  The issues that attend the use of the method for considering
fracture in general will be presented elsewhere, while here we will note
the key elements in carrying out such simulations.  

In order to investigate the interaction between an advancing crack and
a GB, we consider the $\Sigma 21 (\bar{2}41)$
GB in fcc nickel.  A crack is initiated in one of the
grains by removing a single (111) plane, such that the crack tip
is located about $2$ nm from the GB.  The crack
is then loaded by applying the isotropic linear elastic displacement
fields for a sharp crack at the mesh boundaries.  The load is
incrementally increased by scaling the boundary node displacements and
allowing the interior nodes to relax to their minimum energy
configuration.  

\begin{figure}
\narrowtext \epsfxsize=3.4truein \vbox{\hskip 0.15truein
\epsfbox{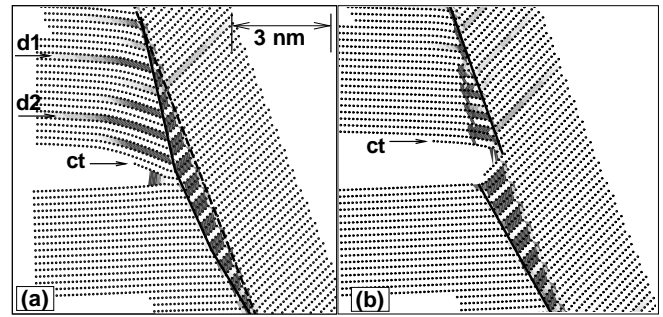}}
\medskip
\caption{Crack/GB interaction.  Snapshots of crack tip
region showing motion of crack tip, dislocations and GB. }
\end{figure}

Two snapshots of the solution are shown in fig. 3.  We show the atoms
associated with that part of our finite element mesh that is fully
refined to the atomic scale in the immediate vicinity of the crack
tip.  The surrounding mesh, which extends about 300 nm in each
direction, has been removed for clarity.  The dots are atomic
positions, while the contours reveal displacement jumps across active
slip planes, indicating the presence of dislocations.  Frame (a) of
this figure is the configuration after 4 load steps. The atom labeled
``ct'' indicates the initial location of the crack tip, and one can
see that the crack has begun to propagate towards the GB by cleavage.
Light grey slip traces emanating from the GB, such as those labeled
``d1'' and ``d2'' show where the stressed GB has emitted
dislocations, some of which have been pinned against the edge of the
fully refined region.  The dashed line running diagonally through the
figure indicates the initial location of the GB which
moves as a result of the high stresses in the crack tip region.  This
motion is accommodated by the structural rearrangement of atoms in the
left hand grain to lattice sites of the right hand grain due to
shearing along atomic planes.  The solid line through the figure
indicates the location of the GB after migration.

In frame (b) of figure 3, the solution after another few load steps is
depicted.  Here, the crack has reached the GB and has been
blunted when atoms above the plane of the crack again underwent a
shearing deformation.  This time, however, the right hand grain shears
to match the structure of the left hand grain.  The two straight solid
lines indicate the new location of the GB.  The result of
this crack blunting is a significant reduction in the stress levels
above the crack, and dislocations such as ``d1'' and ``d2'' in the
first frame have moved back to be re-absorbed by the GB.
Further loading of the crack leads to a continued crack blunting due
to shearing of atomic planes along the GB.  We have
studied other GBs \cite{miller}, where the crack has
deflected and continued to propagate along the GB, in
contrast to the crack blunting mechanism described here.

\begin{figure}
\narrowtext \epsfxsize=3.4truein \vbox{\hskip 0.15truein
\epsfbox{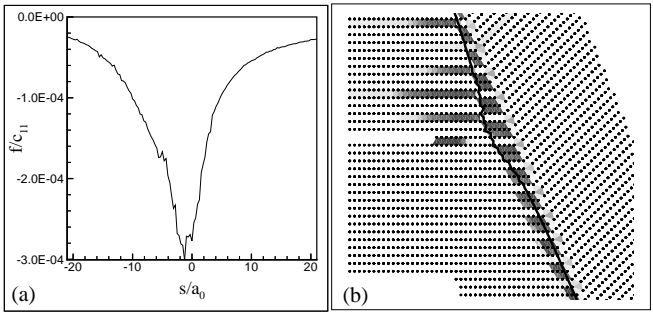}}
\medskip
\caption{(a) Driving force (normalized by the elastic constant,
$c_{11}$ ) as a function of position $s$ along the GB
(normalized by the lattice constant $a_0$). (b) The same force
superimposed on the GB for comparison.}
\end{figure}

To understand these results, we turn to continuum mechanics which
provides the basis for evaluating the energetic origins of GB
migration.  Such reasoning asserts that the driving force on an
interface is given by the jump in the Eshelby tensor \cite{eshelby}
across the interface, with this tensor defined as
\begin{equation} 
P_{ij}=W\delta_{ij}-u_{k,i}\sigma_{kj}.  \non
\end{equation}
$W$ is the strain energy density, $u_{i,k}$ is the $k^{th}$ component
of the gradient in the $i^{th}$ component of displacement and
$\sigma_{kj}$ is the stress tensor.  Within the confines of linear
elasticity, we have computed the driving force on the interface by
using a conventional anisotropic linear elastic constitutive model in
conjunction with the standard finite element method to obtain the fields
associated with the crack/GB geometry described above.
Once these fields are obtained, the resulting driving force may be
obtained by computing the jump in the Eshelby tensor across the
interface.  If we further assume that the GB migration is proportional to
the driving force, the driving force profile may be compared directly
with the bowed out geometry as shown in figure 4.

In this letter, we have shown how our mixed atomistic and continuum
analyses has been adapted to the treatment of interfacial deformation.
Such calculations demanded generalization of the original
quasicontinuum formulation to allow for the existence of more than one
grain at the same time.  As validation of the method, we have computed
the structure and energetics of a series of different GBs and found
entirely satisfactory correspondence between these calculations and
those resulting from direct atomistics.  The method has been applied
to three distinct problems: deformation of a stepped twin boundary, the
interaction between dislocations and a GB, and the propagation of a
crack into a GB.  The calculations on the stepped GB allowed for
quantitative evaluation of the stress to move the GB, while
calculations on the crack/GB interaction revealed stress induced GB
motion which can be rationalized in terms of the driving force on that
interface as implied by the jump in the Eshelby tensor.  The advantage
of the model presented here over standard atomistic calculations is
the significant reduction in the computational effort through careful
reduction of the degrees of freedom.  For example, the number of
degrees of freedom associated with the mesh of figure 2a is about
$10^4$, while the same atomistic calculation would have required more
than $10^7$ degrees of freedom.  This approach allows for the
simultaneous treatment of defects occurring over many length scales,
ranging from individual dislocations to GBs and cracks.

We are grateful to C. Briant, R. Clifton, B. Gerberich, P. Hazzledine,
S. Kumar, D. Rodney and A. Schwartzman for discussions, to S.W. Sloan
for use of his Delaunay triangulation code and to M.  Daw and
S. Foiles for use of their Dynamo code.  We are also grateful to AFOSR
who supported this work under grant number F49620-95-I-0264 and the
NSF through grants CMS-9414648 and DMR-9632524 and the DOE through
grant DE-FG02-95ER14561.  RM acknowledges support of the NSERC.

\end{document}